\documentclass[pre,aps,twocolumn,showpacs,floatfix]{revtex4-1}
\usepackage{graphicx}
\usepackage{dcolumn}
\usepackage{bm}
\usepackage{hyperref}
\usepackage{color}
\usepackage{epsf,epsfig}
\usepackage{color}
\usepackage{enumerate}

\begin{document}
\title{How human-derived brain organoids are built differently from
  brain organoids derived from 
  genetically-close relatives: A multi-scale hypothesis}
\author{Tao Zhang}
\thanks{These two authors contributed equally.}
\affiliation{Department of Polymer Science and Engineering, School of Chemistry and Chemical Engineering, Shanghai Jiao Tong University, Shanghai 200240, China}
\author{Sarthak Gupta}
\thanks{These two authors contributed equally.}
\affiliation{Department of Physics, Syracuse University, Syracuse, NY 13244 USA}
\author{Madeline A. Lancaster}
\affiliation{MRC Laboratory of Molecular Biology, Cambridge Biomedical
  Campus, Francis Crick Avenue, Cambridge CB2 0QH, UK}
\author{J. M. Schwarz}
\affiliation{Department of Physics, Syracuse University, Syracuse, NY 13244 USA}
\affiliation{Indian Creek Farm, Ithaca, NY 14850 USA}
\date{\today}

\begin{abstract}
How genes affect tissue scale organization remains a longstanding
biological puzzle. As experimental efforts are underway to solve this puzzle via
quantification of gene expression, chromatin organization, cellular
structure, and tissue structure, computational modeling efforts remain
far behind.  To help accelerate the computational modeling efforts, 
we demonstrate how a cellular-based model for tissues can be merged
with a model of a cell
nucleus that includes a deformable lamina shell and chromatin to begin
to test multiscale hypotheses linking the chromatin scale and the tissue
scale. To be concrete, we turn to an {\it in vitro} system for the
brain known as a brain organoid. We provide a multiscale
hypothesis to distinguish structural differences between brain
organoids built from induced-pluripotent human stem cells and from 
induced-pluripotent gorilla and chimpanzee stem cells.  Recent
experiments discover that a cell fate transition from
neuroepithelial cells to radial glial cells includes a new intermediate
state that is delayed in human-derived brain organoids as compared to
their genetically-close relatives, which significantly narrows and lengthens the
cells on the apical side~\cite{Lancaster2021}. Additional experiments revealed that the transcription factor ZEB2 plays a major
role in the emergence of this new intermediate state with ZEB2 mRNA
levels peaking at this onset~\cite{Lancaster2021}.
We postulate
that the enhancement of ZEB2 expression driving this intermediate
state is potentially due to chromatin reorganization in response to mechanical deformations of the nucleus. More precisely, there exists a larger critical mechanical strain triggering the
reorganization that is higher for human-derived stem cells than their genetically-close relatives, resulting in 
delay of ZEB2 upregulation. We demonstrate how such hypotheses can begin to be 
  computationally tested by exploring how slightly different initial
  configurations of chromatin reorganize in response to applied strain with increasingly different initial configurations representing less genetically-close relatives. We
  find that as the difference in the initial chromatin configuration
  increases the average of the difference in the magnitude of chromatin displacement in response to an applied strain increases faster than
  linearly but more slowly than exponential. We also show how changes in chromatin strain and contact maps in response to applied strain can yield differences between genetically-close species to provide key chromatin organization information in understanding how one species differs in structure from another.  
\end{abstract}
\maketitle
\section{Introduction}
\begin{figure}[h]
	\centering
	\includegraphics[width=0.99\columnwidth]{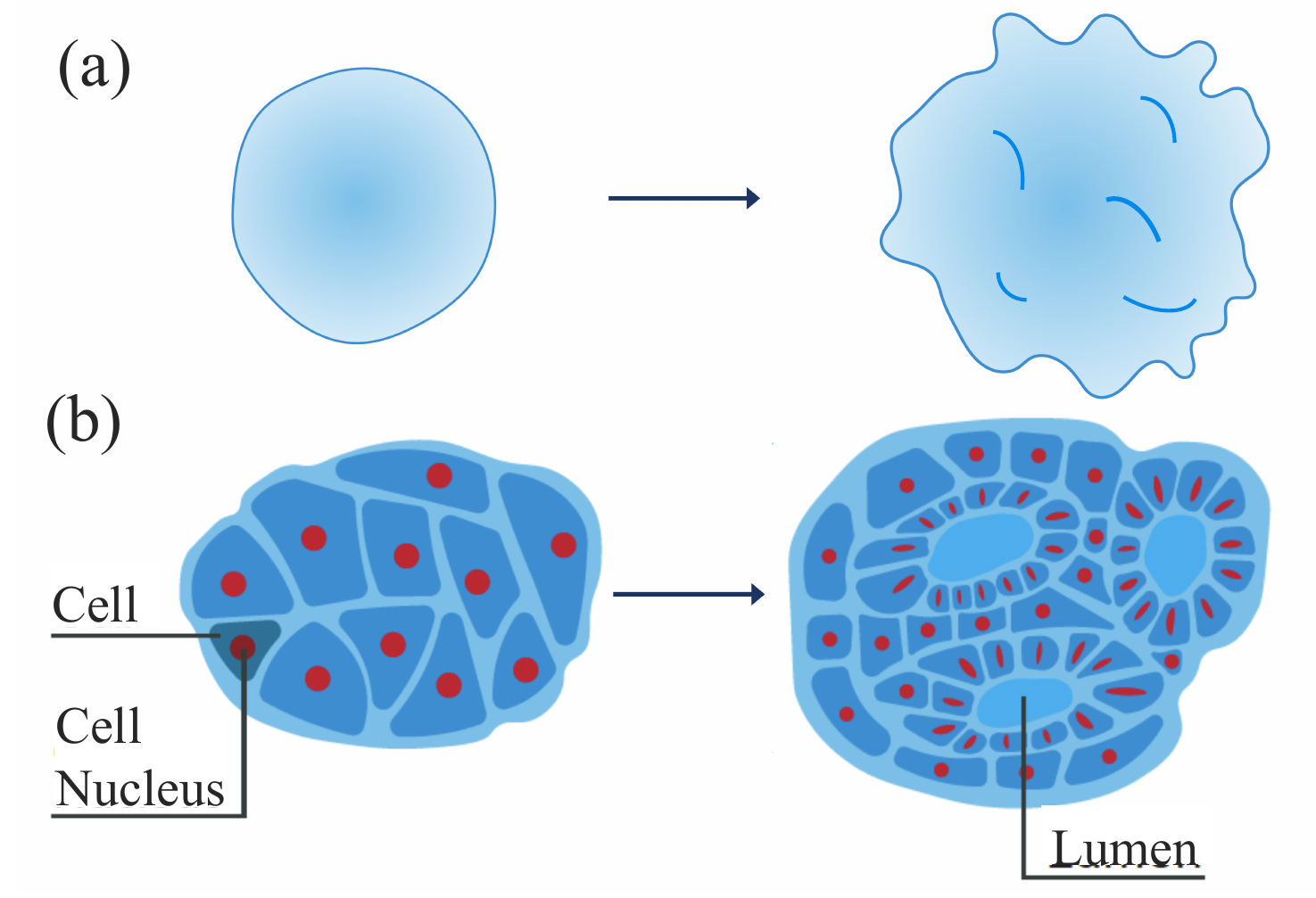}
	\caption{{\it Brain organoid schematics} (a) Schematic of
            a brain organoid at an earlier time and at a later time. (b) Schematic of a
          cross-section of (a) with more cellular detail.}
	\label{fig:brain_organoid}
\end{figure}

Genetic mutations can indeed impact tissue scale organization. For
instance, there is plentiful experimental evidence that changes in gene expression can affect the foliated structure of a developing
brain~\cite{Chenn2002,Chenn2003,Nonaka2013,Bernabe2016}. To be even more specific, mutations of the LIS1 gene result
in lissencephaly, or a smooth brain~\cite{Stahl2013,Reiner1993}. Although the connection between genes and tissue-scale organization is highly complex, with many puzzle pieces still unknown, experimental efforts are actively working to uncover them. Specifically, recent experimental progress on linking the
chromatin scale with the tissue scale is now emerging with, for example, the finding
that mechanical straining a tissue leads to the loss of
heterochromatin to give rise to cell nuclear
softening~\cite{Nava2020}. Structural measurements at the cell and
tissue scale have long been standard in biology. Measurements of the spatial organization of
chromatin in cells using chromatin conformation capture
techniques are now also  
well-established~\cite{Lieberman2009,Rao2014,Wang2016,Nagano2013,Ramani2017,Bonev2016,
Dekker2017,Belaghzal2021}. Other
methodologies acquiring additional information about chromatin
architecture include
immunoGAM~\cite{WinickNg2021} and spatially resolving
chromatin modifications~\cite{Deng2022} will also help put together
this puzzle.  Given these experimental developments, one wonders how
the current state of computational modeling can also help solve this
puzzle.  This manuscript gives a roadmap on how to begin to build
minimal, multiscale computational models to help solve this puzzle.

Given some recent, 
intriguing experimental results on brain
organoids ~\cite{Lancaster2021}, we will use these results as a
guide. As the brain is being
built, it is composed of living, multiscale matter capable of emergent forms of 
mechanical, chemical, and electrical functionality at the genome
scale, the cell nucleus scale, the cellular scale, and/or 
the tissue scale in a nested structure with interplay between the
different scales. While such a multiscale materials neuroscience
viewpoint may seem obvious-but-unwieldy to many, given the theoretical and experimental techniques that have
evolved over the past few decades, we are now on the cusp of being
able to develop quantitative predictions based on this viewpoint for the brain
structure-function relationship, and, more importantly, to test them
using brain organoids~\cite{Lancaster2013,Lancaster2014,Lancaster2014b}---an {\it in vitro} realization of a
developing brain. See Figure 1.

Delving into the experimental findings, even before the onset of neurogenesis, the human forebrain, consisting
of precursor cells known as neuroepithelial (NE) cells, is larger than
other mammals~\cite{Rakic2007}. It has, therefore, been long hypothesized that differences in these NE cells may result in expansion of the
neocortical primordium~\cite{Rakic1988,Rakic1995}. The expansion begins as tangential expansion
and then becomes radial as asymmetric NE cell division emerges with
one daughter radial glial (RG) cell (and the other daughter a NE cell)
~\cite{Gotz2005}. The RG cells do not inherit epithelial features of NE cells
and are rather elongated and, presumably, provide patterning for
neurons. Since it is difficult to explore this hypothesis in humans
and apes, recent experiments study human and ape brain organoids
derived from induced pluripotent stem cells (iPSCs)~\cite{Lancaster2021}. Intriguingly, human-derived brain organoids exhibit larger
surface area than their ape-derived counterparts~\cite{Lancaster2021}. In studying the
NE-RG cell transition in such brain organoids, an intermediate cell
morphology was discovered and named transitioning NE (tNE) cells. In
tNE cells, cell shape changes occur prior to the change in cell
identity.  This intermediate cell morphology is delayed in human brain
organoids in comparison with ape brain organoids.  Since the delay in
tNE formation postpones the transition from tangential-to-radial
expansion, this delay, combined with a shorter cell cycle for human
progenitor cells, leads to a larger progenitor pool and, thus,
typically larger human-derived brain organoids. 

\begin{figure*}[t]
	\centering
	\includegraphics[width=0.99\textwidth]{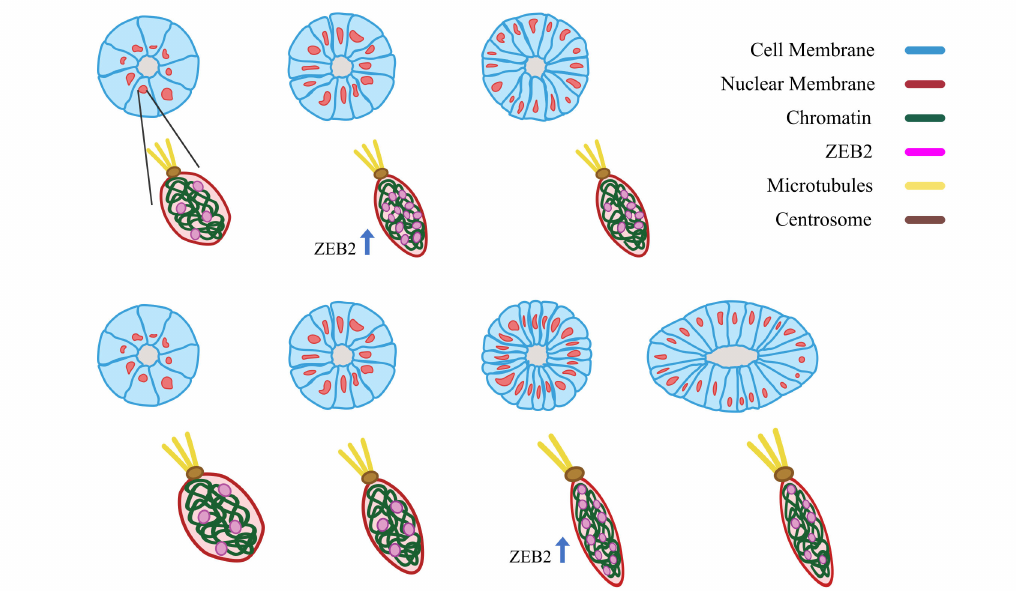}
	\caption{{\it A multiscale hypothesis}: The top row is a
          sketch of the time evolution of the cross-section of one
          cortex-lumen structure in a gorilla-derived brain organoid.  The
          bottom row depicts the cross-section of the same structure,
          but in a human-derived brain organoid. According to our 
          multiscale hypothesis, there is a 
          critical strain on a cell nucleus to help initiate
          upregulation of ZEB2.  The critical strain is larger for human-derived
          pluripotent stem cells as compared to gorilla-, or
          chimpanzee-derived pluripotent stem cells, resulting in a
          delay in the human-derived brain organoids. Note that ZEB2
          can take on multiple roles, including inhibiting BMP-SMAD
          signaling by inhibiting BMP4 transcription to disrupt
          cell-cell junction formation and regulating the production of
          the microtubule-centrosome binding protein ninein. Note that that all components, particularly ZEB2, are not necessarily to scale. Graphics
          credit: Savana Swoger.}
	\label{fig:multiscale_hypothesis}
\end{figure*}

To understand the molecular mechanisms behind this delay in tNE
formation in human NE cells, time-resolved sequencing analysis helped
to identify differential expression in the zinc-finger transcription
factor ZEB2 and, so, a potential driver of tNE cell formation~\cite{Lancaster2021}.  ZEB2
as a driver was then tested in mutant ZEB2+/- brain organoids as well
as controlling ZEB2 expression such that the human-derived and
ape-derived brain organoids achieve a similar size and
morphology with, for example, the addition of doxycycline to induce
ZEB2 expression at earlier stages.  Additional treatments validated this hypothesis. 

Given such findings, we now ask how does ZEB2, and potentially other
players, regulate the delay in the NE-RG cell transition? To answer
this question requires understanding of what lies in a cell
nucleus. Transcription factors are proteins that control the rate of
transcription of genetic information. Of course, these proteins
themselves need to be made and so we must understand what controls
their own expression rates. While there are a number of pathways
regulating transcription given that the NE-RG transition is dominated
by cells elongating and so changing shape, we are going to pursue a
means of regulation that is mechanical in nature----mechanical in that
some initial cell shape change can potentially induce additional cell
shape changes that can potentially lead to chromatin
reorganization. Since genetically-close relatives do not have the same
chromatin organization, particularly at the topologically-associating
domain scale~\cite{Eres2019}, it would behoove us to ask how such
differences in chromatin organization change as the cell nucleus
changes shape.  Do the differences become amplified or not, or remain
the same? And how do such differences modify gene expression? Here, we
pose a multi-scale hypothesis that may provide a mechanism for delaying 
the NE-RG cell transition and introduce a computational model that can
begin to address chromatin reorganization as a function of cell
nucleus shape change.

\section{A multiscale hypothesis}
What do we mean by a mechanical means of regulating transcription? Let
us consider human iPSCs. Genetic information is stored in the cell
nucleus and when combined with histones, forms chromatin. Chromatin is
spatially and temporarily organized within the cell nucleus.  While a
difference in genetic sequence between, say, a chimpanzee and a human,
is rather small----approximately about 1.2\%~\cite{Ebersberger2002}----perhaps even this rather
small difference in genetic sequence translates into differences in spatial organization of the genome inside a cell
nucleus. Incidentally, Hi-C maps of human versus chimpanzee stem cells 
demonstrate differences~\cite{Eres2019}. Such differences in spatial organization of the
genome can potentially translate into differences in gene expression
dynamics, such as ZEB2. Moreover, the spatial organization of
chromatin can be modified by a change in the shape of a nucleus, which
is often due to a change in the shape of the cell with cell nuclear
shape often mimicking cell shape~\cite{Versaevel2012,Lele2018}.  As evidence for this,
Golloshi, {\it et al.}, study chromosome organization before and after
melanoma cells travel through 12 micron and 5 micron constrictions to
find compartment switching between euchromatin and heterochromatin,
among other differences, when performing the Hi-C analysis~\cite{Golloshi2022}.

As the NE cells divide, given the brain organoid is developing in a
confined environment, we hypothesize that the additional cells
generate compression on, say, a cell of focus.  As the compression
increases, there is presumably a slight change in cell shape, which
may result in a change in nuclear shape, which then may result in a
change in the spatial organization of the chromatin. For instance, a
slight compression in a particular direction (and hence elongation of the nucleus in the
direction perpendicular to the compression), may open up a chromatin
region to facilitate/enhance ZEB2 expression.  With this enhancement,
presumably ZEB2 is able to take on additional functionality. See Figure 2. 

 {\it We hypothesize that the
amount of compression and/or compression rate required to modify the
chromatin organization associated with ZEB2 expression 
varies from human iPSCs to ape iPSCs. More precisely, a higher amount
of compression is needed for human iPSCs as compared to gorilla- or
chimp-derived iPSCs. }

Once ZEB2 expression increases, there are multiple downstream effects
that can impact cell shape. For instance, 
ZEB2 is a regulator of SMAD signaling that can affect the production
of cell-cell
junction proteins~\cite{Vandewalle2005}. Should the upregulation of ZEB2 lead to
fewer cell-cell junction proteins at the apical side, then the cells are able to more
readily contract at the apical side. Moreover, the actin-binding protein
SHROOM3 helps strengthen the stress fibers oriented in such a way to
facilitate constriction~\cite{Chu2013}. Manipulation of SMAD signalling
resulted in influencing the onset of tNE morphology, while treatment
with LPA countered the apical constriction~\cite{Lancaster2021}. In addition to
diminishing the strength of cell-cell junctions, enhancing apical constriction,
microtubule organization may also be affected. It is known that ZEB2
regulates the production of the microtubule-centrosome binding protein
Ninein~\cite{Srivatsa2015}, which may help guide the cell fate transition towards
a RG cell given that microtubules shape RG morphology~\cite{Nulty2015}. Moreover, Fouani, {\it et al.} find that ZEB1 switches
from being a transcription factor to a microtubule-associated protein
during mitosis~\cite{Fouani2020}.  And while they do not find the same phenomenon for
ZEB2, the multi-functionality for this class of proteins is rather
intriguing~\cite{Birkhoff2021}.  In any event, given the downstream changes to cell
adhesion and cell cytoskeletal organization to alter the cellular
forces at play, the cell
shape transition to more
elongated cells drives radial-like expansion of the brain organoid.


Our chromatin-reorganization-due-to-cell-compression hypothesis is readily testable using Hi-C at the single iPSC level to
determine at what amount of compressive strain does, or does not, alter
the chromatin organization pertaining to ZEB2 expression. At the
brain organoid level, the mechanical perturbations are self-generated,
if you will, by the cells and the influence of the environment in which
the brain organoid is embedded. The more cell nuclei become deformed,
the more we need their explicit description in cellular-based
models. Note that we would like to go beyond the typical biochemical
signaling pathways through cell-cell junctions, focal adhesions, or
YAP/TAZ by which others have studied nuclear
mechanotransduction~\cite{Athirasala2017} to explore directly chromatin organization.

Experimental tests of this multi-scale phenomenon will either
validate, or not validate, the hypothesis.  Here, we ask the question:
How can we build minimal, multi-scale models to computationally
generate such hypotheses prior to performing gene expression
experiments such that the modeling informs the experiments as opposed
to experiments informing the modeling? We argue that several of the
multi-scale pieces are coming into focus to allow us to more readily
connect genetic-scale processes to tissue-scale processes, though
there is still much to do.  The pieces that are coming into focus are
cellular-based models for the structure of organoids as well as
structural models of deformable cell nuclei containing chromatin ~\cite{Zhang2022,Liu2021}.

\begin{figure*}
	\centering
	\includegraphics[width=0.99\textwidth]{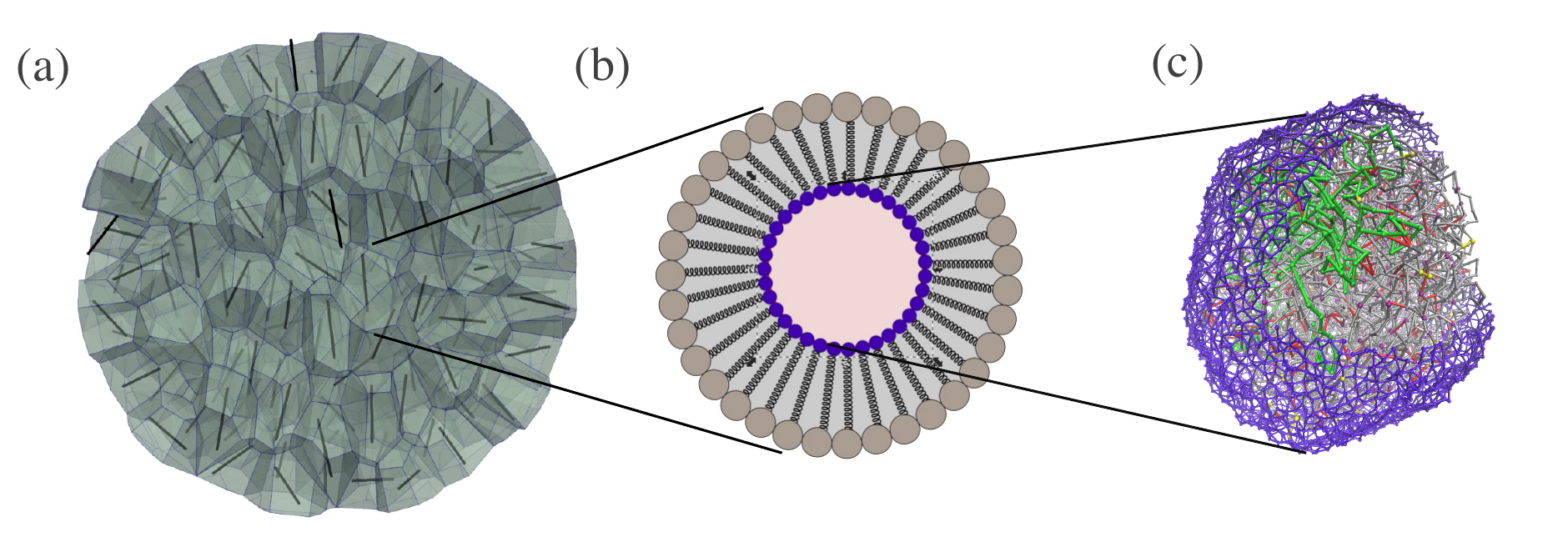}
	\caption{{\it A computational approach to multiscale modeling
            of tissue structure}: (a) A representative organoid based
          on a three-dimensional vertex model in which cells are
          represented as deformable polyhedrons and there are no gaps
          between the cells. The black rods denote the long axis of
          the polyhedron as determined by a fit to a minimal volume
          ellipsoid. (b) A schematic of a two-dimensional
          cross-section of a cell that includes the acto-myosin cortex
          (outer ring of springs), the lamina shell (inner ring of
          springs), and the bulk cytoskeleton, including
          vimentin (the springs connecting the inner and outer ring of
          springs). (c) A deformable lamina shell cell nuclei (purple)
        containing chromatin (grey). A portion of the chromatin is
        colored in green to highlight its configuration. }
	\label{fig:multiscale_modeling}

\end{figure*}

We developed a three-dimensional vertex model to predict the early
development morphology and rheology of cellular
collectives~\cite{Zhang2022}, representing cells as deformable
polyhedrons without gaps, termed confluent. Our model, applied with
periodic boundary conditions, reveals a rigidity transition at a
critical shape index of \(s_0^* = 5.39 \pm 0.01\). We observed a
distinct boundary-bulk effect in confluent collectives subjected to
lateral and radial extensile deformations, with the effect spanning a
single-cell layer thickness. Specifically, cells within the bulk align
less with lateral deformations compared to boundary cells, indicating
that internal cells are largely shielded from deformations over slow
timescales and moderate strain. This finding sheds light on cell shape
patterning mechanisms in organoids and live organisms. We also developed a computational model at the cell nucleus scale to
quantify nuclear mechanics and morphology~\cite{banigan2017,Liu2021,Berg2023}. The model incorporates activity as
well as the deformability of the lamina shell, chromatin-lamina
linkages, and chromatin crosslinks. In addition to quantitatively capturing
stress-strain curves revealed in experiments, it also provides a new
mechanisms for correlated chromatin motion and for nuclear bleb initiation~\cite{banigan2017,Liu2021,Berg2023}.

In the Computational Model Section, we detail these two recent
computational models: one focused on the tissue scale and the other examining the cell nucleus scale. We will then demonstrate how they can be
coupled to continue to probe a key brain organoid structure question
that has been recently asked and begun to be answered experimentally: How does the
development of the structure of human-derived brain organoids differ from their
closest genetic relatives, namely chimpanzee-derived and
gorilla-derived brain organoids? We will demonstrate,
in principle, how our multiscale hypothesis can be first tested computationally to determine its feasibility. In other words, we provide a roadmap for a direction of next-generation multi-scale, cellular-based computational
models that are minimal---minimal in the sense that complexity emerges
from simplicity as opposed to complexity emerging from complexity. They can also be falsified by experiments as they
yield predictions about organoid shape and rheology, cell shape and
rheology and cell nucleus shape and rheology and chromatin structure,
ultimately. With falsification, comes progress of some form. See Figure 3 for an overview of the computational
framework that we now detail.

\section{Computational Model} 

\subsection{A cellular-based computational model for organoids} 

Let us begin with a   
cellular-based, computational model for an organoid that is rooted in
earlier
work~\cite{Honda1982,Honda2004,Farhadifar2007a,Staple2010a,Okuda2013,Okuda2015,
Bi2015,Sahu2020a}. With such a model, we
can track changes in cell shape, which can potentially give rise to
changes in nuclear shape such that changes in nuclear shape can
potentially lead to changes in chromatin organization. To this end, let us review recent
construction of a what is called a three-dimensional vertex model with
boundaries~\cite{Zhang2022}. 

Cells are biomechanical and biochemical systems that operate out of equilibrium, driven by internal or active forces. The biomechanics of an organoid, which is a cluster of cells, is described by the energy functional:
\begin{equation}
\label{eq:energy}
E= K_V\sum_{j}(V_{j} - V_{0})^2+K_A\sum_{j}(A_{j} - A_{0})^2 +\gamma\sum_{\alpha}\delta_{\alpha,B} A_{\alpha},
\end{equation}
where $A_{j}$ is the total area of the $j$th cell, $V_{j}$ is its volume, and $\alpha$ represents the faces of the cells. The term $\delta_{\alpha,B}$ is 0 if a cell face is not at the boundary B of the collective and 1 otherwise. The formula includes penalties for deviations from a cell's preferred volume and area, with $K_V$ and $K_A$ being the stiffness coefficients for volume and area, respectively. The volume term reflects the cell's bulk elasticity, with $V_0$ as the target volume.

The area term, when expanded, includes quadratic, linear, and constant parts. It is typically suggested that the quadratic term relates to the contractility of the actomyosin cortex, while the linear term, with a coefficient of $-2K_A A_0$, results from the balance between cell-cell adhesion and cortical contractility. Negative values, representing smaller $A_0$, indicate dominance of cortical contractility, and positive values, representing larger $A_0$, indicate dominance of cell-cell adhesion.

Indeed, cell-cell adhesion and contractility are coupled~\cite{Hoffman2015}. For
      instance, knocking out E-cadherin in keratinocytes, effectively
      changes the contractility ~\cite{Sahu2020a}. Given this intricate coupling,
      it may be difficult to tease out the competition. Moreover, the
      finding in two-dimensional vertex models of a rigidity
      transition as the target perimeter is increased then leads to
      the interpretation that unjamming, or fluidity, is given by an
      increase in cell-cell adhesion, which appears to be 
      counterintuitive~\cite{Bi2015}. As for an alternative
      interpretation, by adding a constant
      to the energy, which does not influence the forces, again, the energy can be written in the above quadratic
      form. Since we cannot
      tune cell-cell adhesion independently of cortical contractility,
      we posit that the target area is simply a
      measure of the isotropy of cortical contractility, assuming that
      curvature changes in the cells remain at scales much smaller than the
      inverse of a typical edge length. It is the cell-cell adhesion
      that is bootstrapped to the cortical contractility as cell faces 
      are always shared. To be specific, the larger
      the target area, the less isotropically contractile the cell is,
      and vice versa. The less isotropically contractile a cell is,
      the more it can explore other shapes to be able to move past
      each other in an energy barrier-free manner resulting in fluidity. Additional terms
      linear in the area for specific cell faces complexify the notion
      of isotropic contractility.


Regarding the linear area term in Eq. (1), for cells located at the boundary of the organoid, an additional surface tension term is introduced for faces that interact with the "vacuum," which consists of empty cells. These empty cells do not exert forces on the other cells, but they enable the surface cells of the organoid to relax. Additionally, at this stage, there is a constraint preventing cells from separating from the organoid.

For any length $l$ in the simulation, it can be nondimensionalized using the relation $l = V_0^{1/3}$. This process highlights a crucial parameter in these models, the dimensionless shape index $s_{0} = A_{0}/(V_{0}^{2/3})$, which correlates with the target area. For instance, a regular tetrahedron possesses a dimensionless shape index of approximately $s_0 \approx 7.2$.

We have addressed the biomechanical aspect of cells with biochemical
aspects indirectly encoded into the model parameters. We must
also account for their dynamics. Cells can move past each other even
when there are no gaps between them. In two dimensions, such movements are
known as T1 events. Understanding these events are key to understanding
the rigidity transition in two dimensions~\cite{Bi2015}.   In three dimensions, such
movements are known as reconnection events.   Prior work has developed
an algorithm for such reconnection events focusing on edges becoming
triangles and vice versa that may occur for edges below a threshold
length $l_{th}$ for a fixed topology~\cite{Okuda2013}. Specifically,
each vertex has four neighboring vertices and shares four neighbor
cells. Each edge shares three cells and each face/polygon shares two cells. Our modeling builds on that key work~\cite{Okuda2013}. 

In addition to reconnection events, there is an underlying Brownian dynamics
for each vertex. Specfically, the equation of motion for the position ${\bf r}_I$ of a single vertex $I$ 
is 
\begin{equation}
\dot{\mathbf{r}}_{I}=\mu \mathbf{F}_{I} +\mu \mathbf{F}_{I}^B,\label{eq:dyn} 
\end{equation}
with $\mathbf{F}_{I}$ and $\mathbf{F}_{I}^B$ denoting the conservative force and
the random force due to active fluctuations on the $Ith$ vertex respectively. 
The force $\mathbf{F}_I$ is determined from both the area and volume
energetic constraints and, hence, includes cell-cell
interactions. Moreover, each vertex performs a random walk with an effective
diffusion coefficient of $\mu k_B T$, where $T$ is an effective
temperature. Unless otherwise specified, the mobility $\mu=1$.
Finally, the Euler-Maruyama integration method is used to update the
position of each vertex.

\subsection{A computational model of deformable cell nuclei} 

The cell nucleus houses the genome, or the material containing
instructions for building the proteins that a cell needs to
function. For humans and other genetically-close relatives, this
material is $\sim 1$ meter of DNA. Using proteins to form chromatin,
the DNA 
is packaged across multiple spatial scales to fit inside an $\sim 10\
\mu\mathrm{m}$ nucleus~\cite{gibcus13}. In addition, chromatin is highly dynamic; for instance, correlated motion of micron-scale genomic regions over
timescales of tens of seconds has been observed in mammalian cell nuclei~\cite{zidovska13, shaban18, saintillan18, barth20, shaban20}. This correlated motion diminishes both in the absence
of ATP, the fuel for many molecular motors, and by inhibition of the transcription
motor RNA polymerase II, suggesting that motor 
activity plays a key role~\cite{zidovska13,shaban18}. These dynamics
occur within the confinement of the cell nucleus, which is enclosed by
a double membrane and 10-30-nm thick filamentous layer of lamin
intermediate filaments to form a lamina shell~\cite{shimi15, mahamid16,
  turgay17}. The lamina shell is deformable and, as such, one can quantify its
shape fluctuations. Specifically, depletion of ATP diminishes the magnitude of the shape
fluctuations, as does the inhibition of RNA polymerase
II transcription activity~\cite{chu17}.

\begin{figure*}[t]
	\centering
	\includegraphics[width=0.99\textwidth]{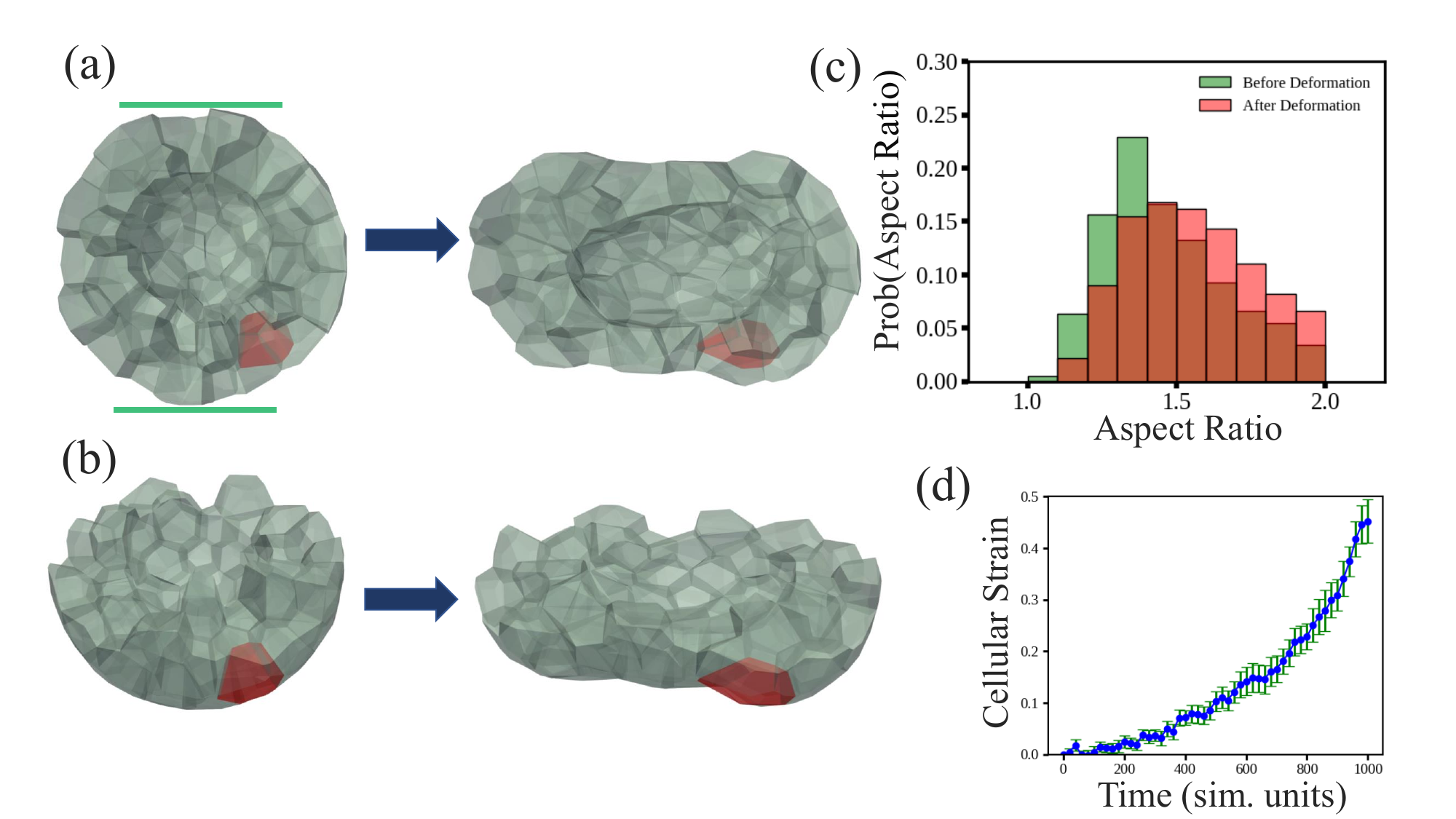}
	\caption{{\it Localized compression of a model brain organoid containing a lumen}:
          (a) Top view of before and after compression a fluid-like
          brain organoid with $s_0=5.6$ and cut in half to better
          expose the interior. The compression occurs in the
          region denoted by the vertical green lines. (b) Side view of
        (a). (c) Probability distribution of the aspect ratio of each
        cell before and after the applied compression/deformation. (d) Plot of the
   cell undergoing the largest change in strain as the organoid
   undergoes compression, averaging over 20 realizations. The corresponding cell is labelled in red in
 (a). }
	\label{fig:organoid_compression}
\end{figure*}

Chromatin and the 
lamina shell interact directly via lamina-associated domains
(LADs)~\cite{guelen08, vansteensel17} and indirectly through various
proteins \cite{dechat08, solovei13,leeuw18}. Therefore, the
spatiotemporal properties of chromatin can potentially influence shape
of the lamina shell and vice versa as the two components are
coupled. Indeed, studies have found that depleting linkages between chromatin and the nuclear lamina, or membrane, results in more
deformable nuclei~\cite{guilluy14, schreiner15}, enhanced curvature fluctuations~\cite{lionetti20}, and/or abnormal nuclear
shapes \cite{stephens19}.  Another recent study suggests that
inhibiting motor activity diminishes nuclear bleb formation~\cite{Berg2022}. Moreover, depletion of lamin A in several human cell lines leads to increased diffusion of chromatin, suggesting
that chromatin dynamics is also affected by linkages to the lamina~\cite{bronshtein15}. Together, these experiments demonstrate the
critical role of chromatin and its interplay with the lamina shell 
in determining nuclear shape. 

To quantify chromatin dynamics {\it and} nuclear shape, we constructed a chromatin-lamina
system with the chromatin modeled as an {\it active} Rouse chain and the lamina as an elastic, polymeric shell with
linkages between the chain and the shell.  We also included chromatin crosslinks,
which may be a 
consequence of motors forming droplets~\cite{cisse13} and/or complexes~\cite{nagashima19}, as well as
chromatin binding by proteins, such as heterochromatin protein I (HP1)~\cite{erdel20}. Recent 
rheological measurements of the nucleus support the notion of chromatin
crosslinks~\cite{banigan17,stephens17}, as does indirect evidence from
chromosome conformation capture (Hi-C)~\cite{belaghzal19}. 
Unlike previous chain-enclosed-by-a-deformable-shell models~\cite{banigan17, stephens17,
  lionetti20}, our model has motor activity. 
We implemented the simplest type of motor, namely extensile and
contractile monopoles that act non-reciprocally on the chromatin. 

To be even more specific, interphase chromatin is modeled as a Rouse chain
consisting of $N$ monomers
with radius $r_c$ connected by Hookean springs with spring constant $K$. We 
include excluded
volume interactions with a repulsive, soft-core potential between any
two monomers, $ij$, and a distance between their centers denoted as $
|\vec{r}_{ij}|$, as given by
\begin{equation}   
U_{ex}=\frac{1}{2}K_{ex}(|\vec{r}_{ij}|-\sigma _{ij})^2
\end{equation}
for $|\vec{r}_{ij}|<\sigma_{ij}$, where $\sigma
_{ij}=r_{c_{i}}+r_{c_{j}}$, and zero otherwise. We include $N_C$ crosslinks
between chromatin monomers by introducing a spring between different
parts of the chain with the same spring constant as along the chain.
In addition to Gaussian fluctuations, we also allow
for explicit motor activity along the chain.  To do so, we assign some number, $N_m$, of chain
monomers to be active. An active monomer has motor strength $M$ and exerts force ${\bf F}_a$ on monomers within a fixed range. 
Such a force may be attractive or ``contractile,'' drawing in
chain monomers, or alternatively, repulsive or ``extensile,'' pushing
them away. Since motors \textit{in vivo} are dynamic, turning off after some characteristic time, we
include a turnover timescale for the motor monomers
$\tau_m$, after which a motor moves to another position on the
chromatin. 

The lamina is modeled as a layer of $M$
monomers connected by springs with the same radii and spring constants
as the chain monomers and an average coordination number $z\approx
4.5$, as supported
by previous modeling~\cite{banigan17, stephens17, lionetti20} and imaging experiments~\cite{shimi15, mahamid16, turgay17}. 
We modeled the chromatin-lamina linkages as $N_L$ permanent springs with stiffness $K$
between shell monomers and chain monomers (Figure 3(c)).
There is an additional soft-core repulsion between monomers making up
the lamina shell to include excluded volume. 

The system, as is the case for the three-dimensional vertex model, evolves via
Brownian dynamics, obeying the overdamped equation of motion:
\begin{equation}
\xi\dot{\bf{r}}_i = ( {\bf F}_{br} +  {\bf F}_{sp} + {\bf F}_{ex} +
{\bf F_{a}}), 
\end{equation}
where ${\bf F}_{br}$ denotes the Brownian/Gaussian force, ${\bf F}_{sp}$ denotes the harmonic forces due to chain
springs, chromatin crosslink springs, and chromatin-lamina linkage
springs, and ${\bf F}_{ex}$ denotes the force due to excluded
volume.

Finally, we used the same simulation parameters as in
Ref. ~\cite{Berg2023} with a varied number of crosslinks as specified
below, though motor activity was turned off.  To implement the
compression, we applied a boundary condition at the top and bottom of
the nucleus, where we moved the boundary at a specified strain
rate. If a lamina particle is beyond the boundary, it experiences a
spring force towards the boundary such that the lamina surface is
compressed due to the moving boundary. Converting from simulation
units to biophysical units, our simulation runs for 500s with a strain
rate of 1.6um/s to compress the nucleus to 60 $\%$ of its original
size, with 30 $\%$ on each side. Ten realizations are obtained for each
parameter varied.

\subsection{Coupling the organoid model to the cell nucleus model}
Now that we have two of the major players in the process----cells and cell nuclei, let us now
envision how we can embed one into the other.  Earlier work has
embedded nuclei in single cells in two-dimensions to understand cell
motility in confinement~\cite{Gupta2021}. In that work, the cell nuclear cortex is connected to the
cell cortex via springs modeling the remainder of the cellular
cytoskeleton, including vimentin, beyond the cell cortex. See Figure
3(b). The model
is, therefore, more detailed than prior minimal models, while still
remaining foundational in that it reveals a new cell polarity
mechanism regulated by vimentin. Given that a direct embedding of cell nuclei of every cell using
springs, as was done in the two-dimensional confined cell motility
study discussed above, is somewhat detailed even in two dimensions, we will do something
simpler. We will instead make the simplifying assumption that nuclear shape tracks
cell shape such that any changes in cell shape result in changes in
nuclear shape.  For instance, should a cell become elongated, its cell
nucleus will also become elongated with the same strain.

To begin to computationally test the notion that chromatin
organization can change in response to tissue structure in
development, we start with a structure that is reminiscent of a
brain organoid structure after several days in development.  We will
begin with a cortex-lumen structure, which is a band of cells
surrounding a fluid-filled core, otherwise known as a
lumen. We used the same simulation parameters as in
  Ref. ~\cite{Zhang2022}, but with fewer cells---152 cells, due to the
  presence of the lumen.  Over the course of several days, the cells in a brain
organoid have divided. Given the confinement of the environment, the dividing cells
become compressed in various directions.  Here, to study the change in shapes of cells as the
  organoid undergoes compression, we compressed the
  ring-shaped area in the center of the organoid inward, decreasing
  the radius by a maximum strain of 40\% over a time interval of 1000
  simulation units. While our compressive strain is external, for
  organoids confined within Matrix gel and undergoing cell division,
  the compressive strains generated are internal.

\section{Results}

\begin{figure*}[t]
	\centering
	\includegraphics[width=0.99\textwidth]{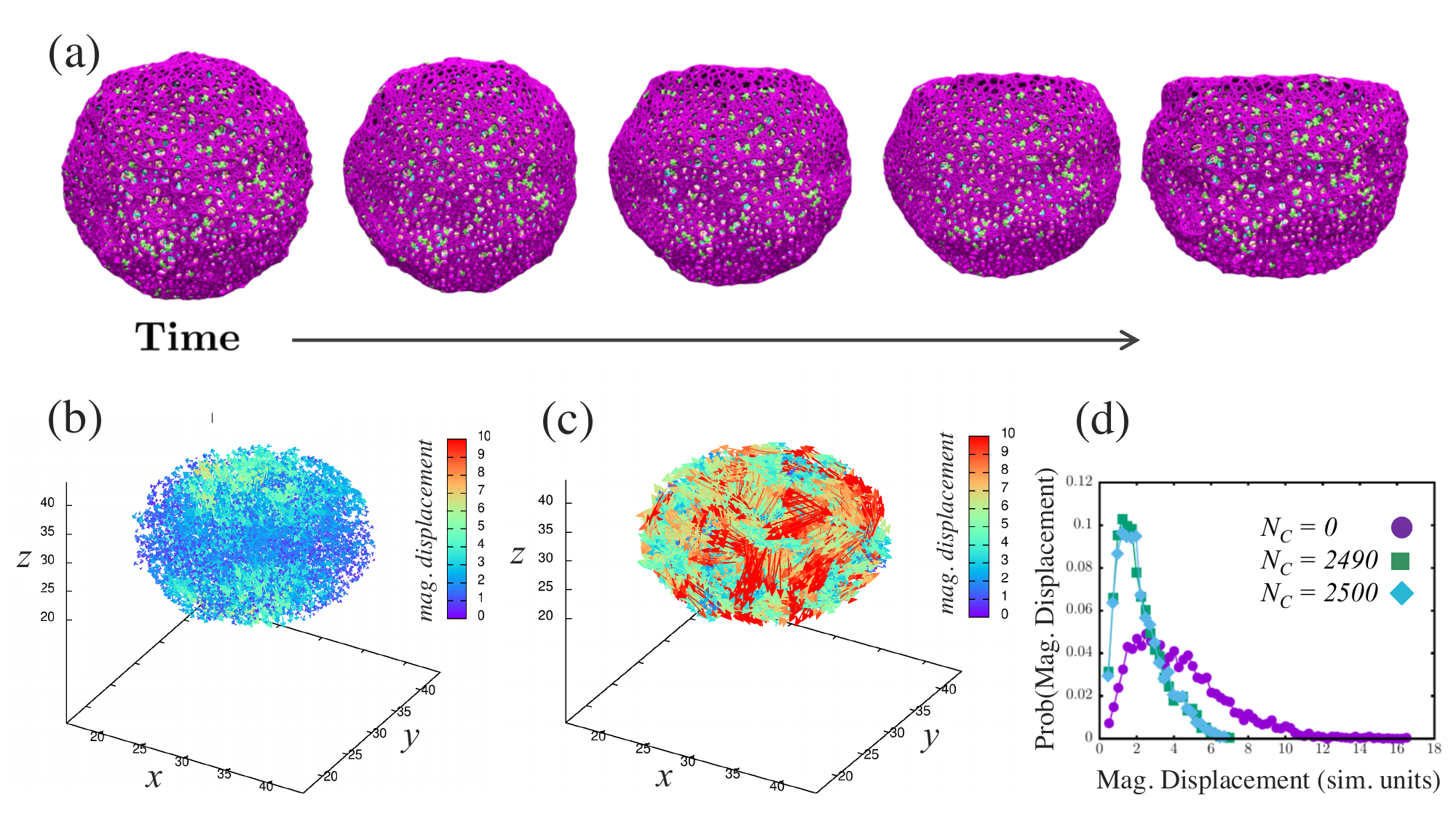}
	\caption{{\it Compressing a model cell nucleus consisting of a
            lamina shell containing chromatin}: (a)
          Snapshots of the model cell nucleus being compressed for
          number of chromatin crosslinks $N_C=2500$ and
          number of linkages $N_L=400$. The strain rate is the same as in Figure 4(d). (b)
          Plot of the chromatin
          displacement field for (a). (c) Plot of the chromatin displacement field
          for $N_C=N_L=0$ for comparison. (d) Probability distribution of the magnitude for
          the chromatin displacement field comparing (b) with (c) and
          also for $N_C=2490$, i.e., a small change from the
          configuration in (a). }
	\label{fig:chromatin_compression}
\end{figure*}

We apply external compression to our model brain organoid, as
discussed in Sec. III. We do this for organoids with a target shape index
of $s_0=5.6$.  See Figure 4. We fit each cell to a minimal volume
ellipsoid and determine its long axis.  In Figures 4(a) and (b), the cell colored red denotes
the cell with the largest change in strain along its long axis. In
Figures 4(c) and (d), we plot the probability distribution of the
aspect ratio of cells before and after the deformation. As a result of
the deformation, the probability distribution becomes more biased
toward cells with higher aspect ratios. Note
that the organoid is not confined in the direction orthogonal to the
deformation and so not all cells will be more elongated. Since applied
strain
is what one can control in the simulations, as opposed to shape, we
also plot the cellular strain as a function of simulation time for the
cell in red in Figures 4(a) and (b). This cell is the cell undergoing
the maximal strain. We will use this information to
study how cell nuclear shape changes.

 \begin{figure*}[t]
	\centering
	\includegraphics[width=0.99\textwidth]{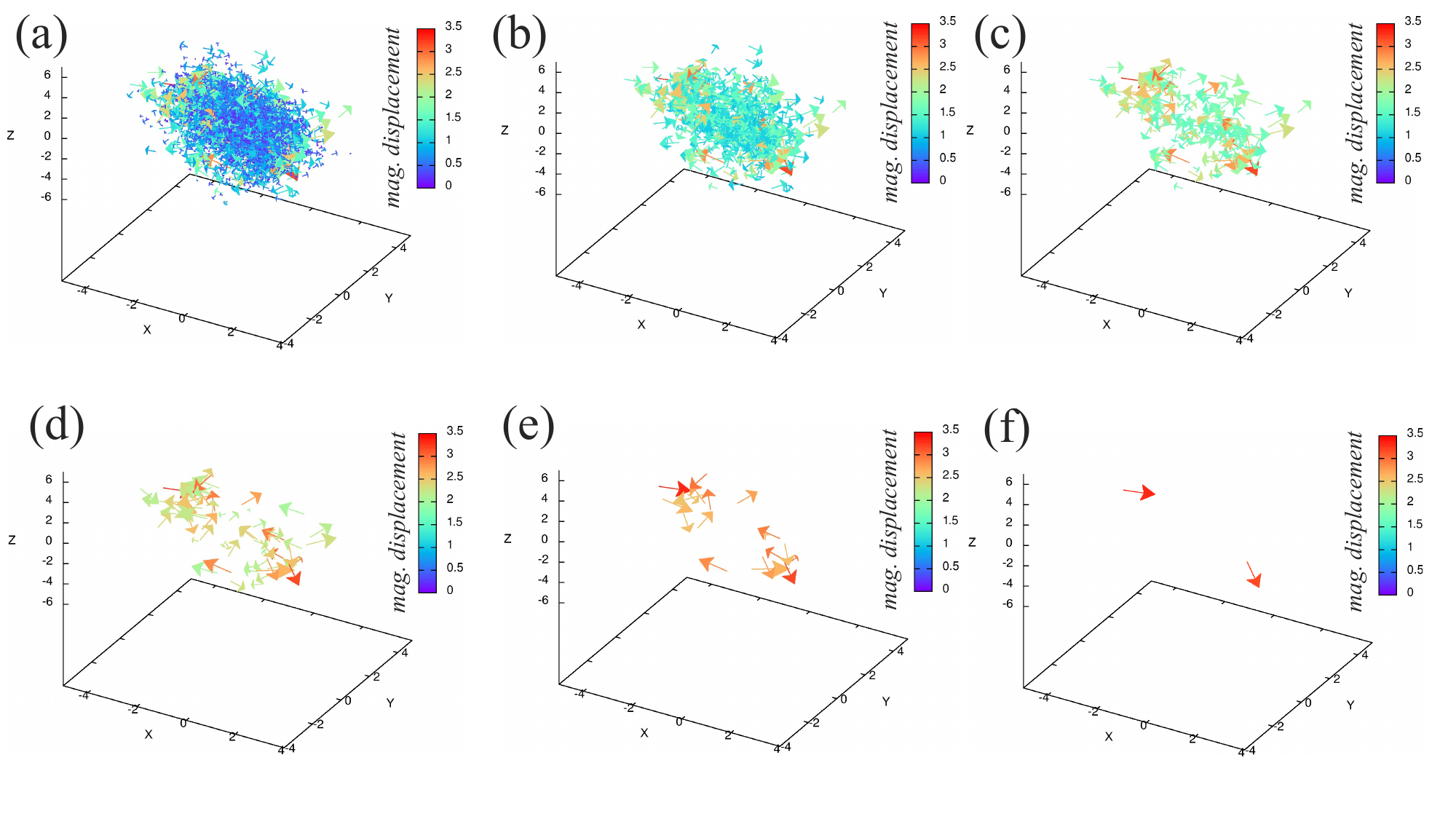}
	\caption{{\it Comparing differences in chromatin displacement
            fields between the unperturbed and perturbed chromatin
            configuration with 0.4\% change in the number of chromatin
          crosslinkers}: (a)-(f) The
          threshold magnitude of the difference in chromatin
          displacements between $N_C=2500$ and $N_C=2490$ 
          increases from top left to bottom right. By increasing the
          threshold magnitude, one can more readily identify regions
          of differences. These regions of differences do not
          necessarily correspondence to the differences in locations
          of the chromatin crosslinkers. }
	\label{fig:differences_displacement_maps}
\end{figure*}

Now that we have the relationship between changes in cell strain,
which is related to cell shape, in response to tissue-scale
compression, we use a simplifying assumption to proceed at the next
smallest length scale. Our simplifying assumption is that cell nuclear
shape tracks cell shape. We then take mechanical deformations that
track cell deformations and apply them to our deformable cell nucleus
to ask how does the chromatin organization change? While our model for
chromatin is a coarse-grained one, we can still ask the following
questions for several types of perturbations: For a given number of
chromatin crosslinks and linkages, by how much does the chromatin
chain locally displace in the presence of uni-axial compression? And
should we perturb the number of crosslinks or linkages by how much
does the local displacements change? We will use small differences in
the number of crosslinkers as a minimal model for slightly different
chromatin structures to represent genetically-close species in which
the pairwise geonomic interactions remain conserved, however, higher
order chromatin structure, such as topologically-associated domains,
appear not be conserved~\cite{Eres2019}. Our generic chromatin model
can teach us something about how the chromatin reorganizes in response to
cell nuclear shape changes given cellular shape changes. 
\begin{figure*}[t]
  \centering
  	\includegraphics[width=0.87\columnwidth]{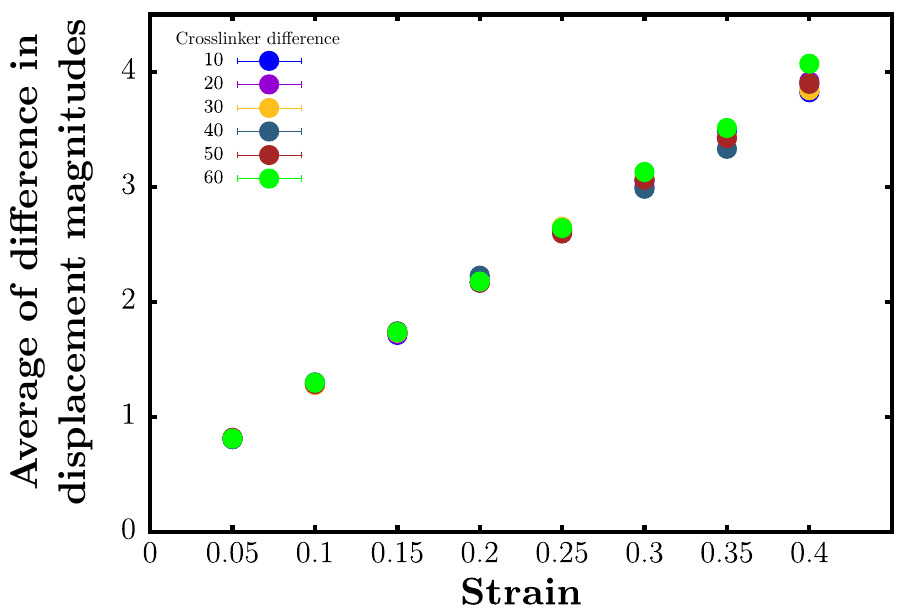}
	\includegraphics[width=0.99\columnwidth]{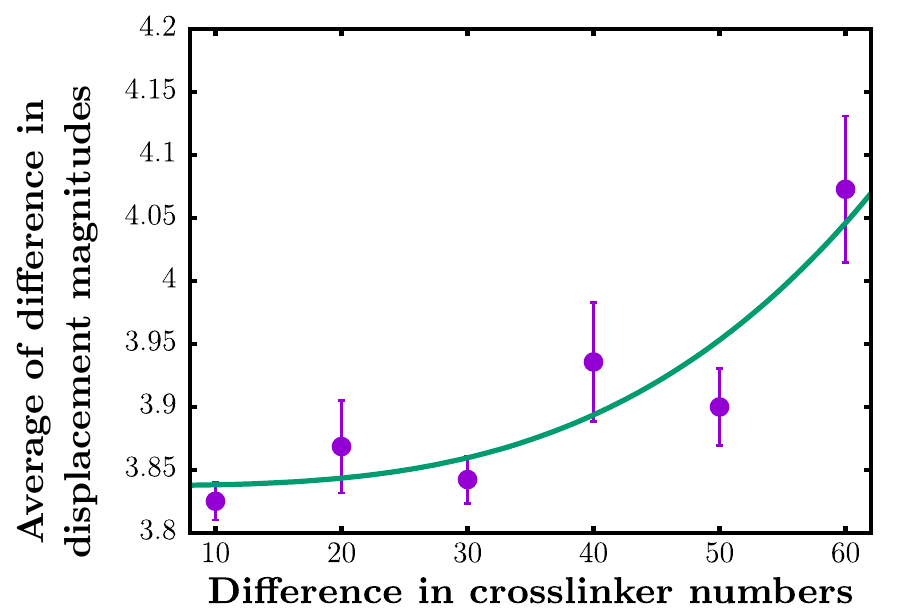}
	\caption{{\it Difference in magnitude of the total average
            chromatin displacement for different numbers of
            crosslinkers for different strains}: Left: Plot of
          the difference in the average of the magnitude of displacements
          between $N_C=2500$ and the number of removed chromatin
          crosslinkers for different values of the strain. Note that
          not until approximately 35\% strain can one distinguish
          between the different numbers of chromatin
          crosslinkers. Right: Plot of the difference in the magnitude of the total average chromatin displacement between $N_C=2500$ and $N_C=2490$ (difference in crosslinker number equals 10) and between $N_C=2500$ and $N_C=2480$, (difference in crosslinker number equals 20), etc., The green curve is the result of a power law fit with an additional constant.  The power law fit yields faster than linear behavior with a power of $\alpha=3.24(5)$. The units of displacement are in terms of chromatin monomer radius.} 
	\label{fig:total_average_displacement_difference}
\end{figure*}

Given the deformation in the maximally deformed cell, as a function of time, we can apply the rate of that deformation to a model cell nucleus and measure the displacement of chromatin monomers, as demonstrated in Figure 5. In Figure 5(a), we show snapshots from uni-axial compression of the lamina shell by two parallel plates moving at constant speed towards each other and exerting force on the lamina shell, but not on the chromatin. We can do so for the same initial chromatin configuration, though with different number of chromatin crosslinks and linkages. Naturally, for a larger number of chromatin crosslinks, one expects smaller displacements. Indeed, we observe smaller displacements for a larger number of chromatin crosslinks with $N_C=2500$ and $N_L=400$, compared with the other extreme with $N_C=N_L=0$, as shown in Figures 5(b) and 5(c). In Figure 5(d) we plot the probability density function of the magnitude of the displacements for the two cases in Figures 5(b) and 5(c), and confirm the fact that the more crosslinked chromatins displaces less. Note that we have only considered one type of perturbation from the reference state here.  For now, we have turned off activity, which could be yet another generator of changes in gene expression for two genetically very similar genomes. 
\begin{figure*}[t]
	\centering
	\includegraphics[width=0.99\textwidth]{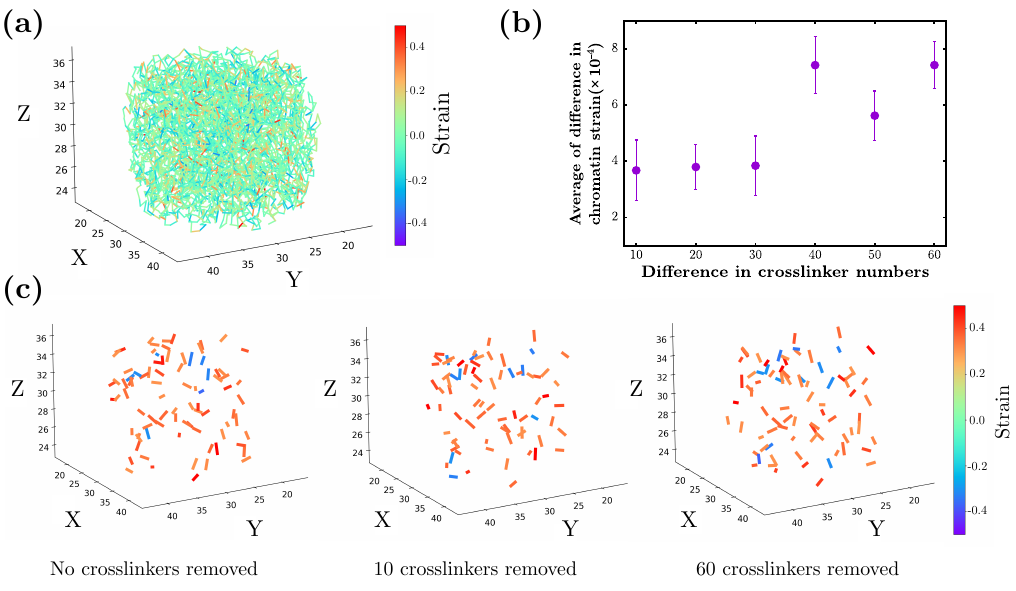}
	\caption{{\it Differences in strains for different numbers of crosslinkers}: (a) Strain map for $N_c=2500$ for one realization at the end of the applied compression. (b) Plot of average of the difference in chromatin strain between $N_C=2500$ and fewer chromatin crosslinks. The green line is a guide to the eye. (c) Stain maps for $N_C=2500$ and for $N_C=2490$ and for $N_C=2440$ with chromatin springs represented only when the magnitude of the strain exceeds 30\%.}
 \label{fig:strain}
\end{figure*}

In what follows, if chromatin were purely a liquid on both short and long spatial scales and time scales, such a comparison between a reference state and a perturbed state in terms of a genetically-close relative under applied compression, as is done here, may not provide much insight. However, experiments demonstrate that the presence of chromatin crosslinks establish the need for a more intricate rheology of chromatin such that it can act elastically over some time and length scales~\cite{Stephens2017,Maeshima2018,Sanders2022}. For instance, mechanical measurements of stretched cell nuclei can be readily explained with chromatin crosslinks~\cite{Stephens2017}. And recent isotropic swelling of cell nuclei reveal very reproducible chromatin configurations before and after the swelling~\cite{Sanders2022}. Other experiments indicate that a Burger's model fits micropipette aspiration data~\cite{Wintner2020}. Moreover, linkages in terms of LADs between chromatin and the lamina shell also potentially provide some elasticity over shorter time scales~\cite{Liu2021}. Presumably, linkages dominate closer to the periphery than in the bulk. Given the complex rheology of chromatin, we perform the following analysis.  

We now determine whether or not a small change in the number of chromatin crosslinks (for the same initial chain configuration) will lead to pockets of differences in displacements in the chromatin. Such pockets could be candidates for changes in genetic expression, even within this minimal model. In Figure 5(d), we plot the probability density function for 10 fewer chromatin crosslinks, for the same initial configuration, and find small differences in the distribution from the unperturbed case. 
We then study the differences in displacements between the largest number of crosslinks and the perturbed case with slightly fewer crosslinks, keeping in mind that with Brownian dynamics, there will always be some difference in the structure.  We are looking for differences above that noise floor. The spatial map of the differences in displacements for one realization (Fig. 6) does exhibit pockets of larger differences
in displacements with such regions being candidates for differences in
gene expression with the assumption that changes in chromatin
configuration can potentially influence genetic regulatory networks at the
base pair level. Those pockets become more pronounced the larger the
threshold magnitude of displacements.

While Figure 6 shows the differences for one
  realization,  we also quantify the difference in the magnitude of
  the total average displacement between the fully crosslinked case
  ($N_C=2500$) and smaller numbers of crosslinks (given the same
  initial configuration of the both the lamina shell and the
  chromatin) for 24 realizations (see Fig. 7). First, we study the
  difference in the magnitude of the total average displacement
  between different species as a function of the amount of applied
  strain (Fig. 7(Left)). We find that for applied strains around 30\%
  or less that the difference in chromatin reorganization, when
  focusing on the larger differences, specifically when the total
  average displacement is above a
  threshold of 3.5 monomer radius, that there is not much difference
  across the species. However, by around 35\% applied strain, a
  statistically significant difference begins to emerge.

  Given the differences across species by 40\% strain, we focus on
  that value of the applied strain and now plot the difference in
  the magnitude in the magnitude of the total average displacement
  versus increasingly different species (see Fig. 7(Right)). We find that as the two
  configurations to be compared are increasingly different numbers of
  crosslinks, that the difference in the magnitude of the total
  average displacement increases. This result is not necessarily
  surprising. However, does this difference diverge exponentially with
  slightly different initial configuration?  If yes, then perhaps the
  system is chaotic.  One would not necessarily anticipate this trend
  as such behavior would presumably hamper the nucleus to carry out
  its functionality in a robust manner. After fitting several
  different phenomenological forms, we find that for $y$ denoting the
  the difference in the magnitude of the total average displacement
  and $x$ denoting the difference in number of crosslinkers,  
  \begin{equation}
    y-b\propto x^\alpha
  \end{equation}
  with $\alpha=3.24(5)$ and $b=3.84(3)$
  reasonably approximates the data. In other words, the difference
  increases faster than linear. Given the complex chromatin rheology, a nonlinear
  scaling is not necessarily surprising. As increasingly different numbers of
  crosslinkers are placeholders for increasingly different species,
  then obtaining such quantities as $\alpha$, even phenomenologically,
  will help further elucidate comparisons between them at the
  chromatin scale. This nonlinear scaling form is a novel type
  of scaling form that should be studied across species. 
\begin{figure*}[t]
	\centering
	\includegraphics[width=0.99\textwidth]{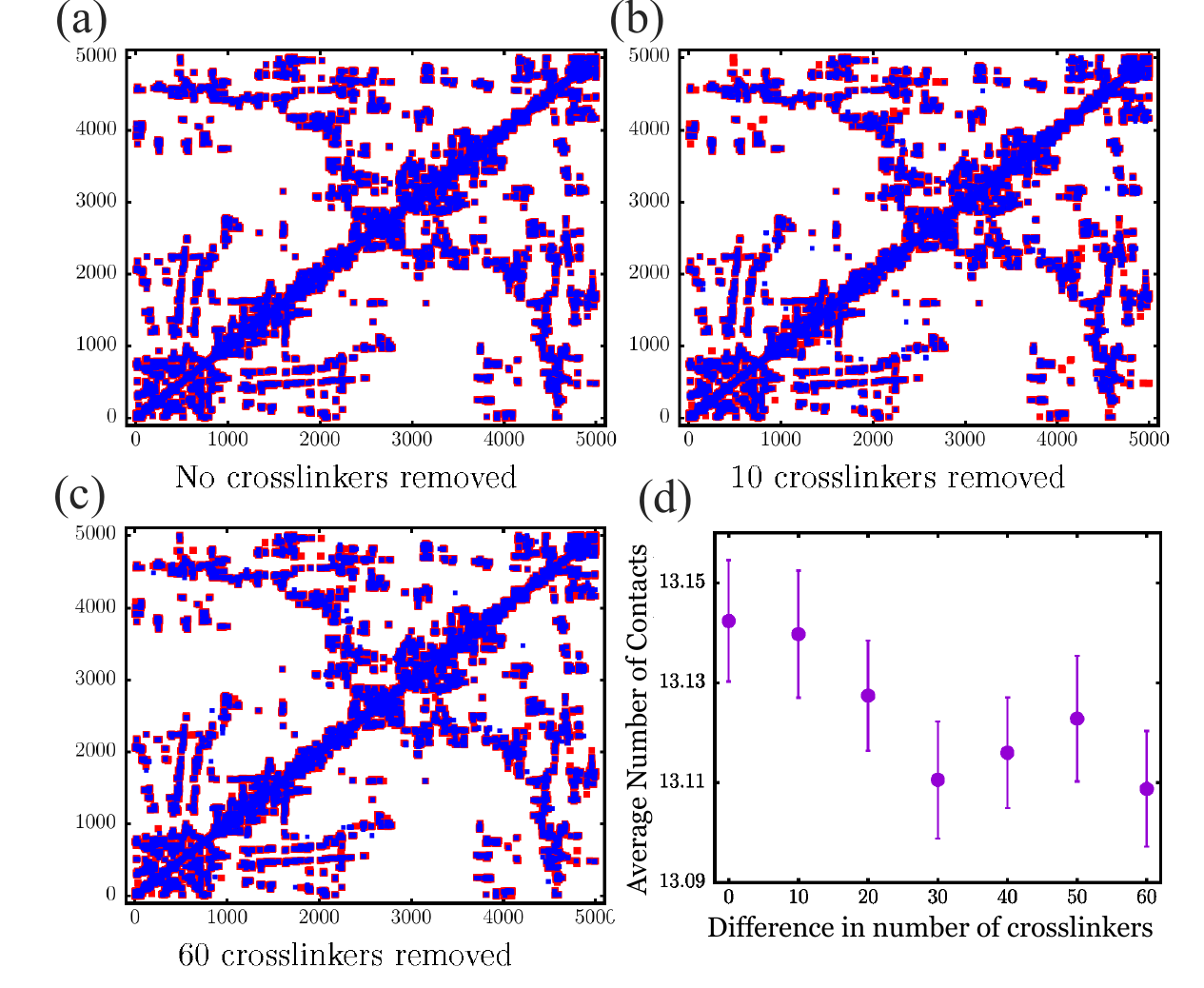}
	\caption{{\it Differences in chromatin contacts for different numbers of crosslinkers}: (a)-(c) Contact maps in which red squares or blue squares represent differences between $N_C=2500$ and $N_C=2490$ and between $N_C=2500$ and $N_C=2440$ and the blue square with red outline represent overlap between the two respective cases. (c) Plot of the average number of contacts per chromatin monomer as the number of crosslinks decreases. }
	\label{fig:contact}
\end{figure*}

We go beyond the differences in displacement of chromatin to also quantify the differences in strain. In Fig. 8(a), the strain map for the spring between each chromatin monomer is plotted for $N_C=2500$ for a typical instance. While most of the springs contain very small amounts of strain, there are springs whose magnitude of strain is greater than 30\% with positive strain, or springs under tension, being favored. Just as in Fig. 7, we also plot the average of the difference in strains between $N_C=2500$ and subsequently smaller numbers of crosslinkers for 24 realizations (see Fig. 8(b)).  This average increases with the increasing difference, though the trend is more difficult to quantify than the average of the difference in magnitude of chromatin monomer displacements. Thus, the green line is merely a guide to the eye. However, as the chromatin configurations become more different given the increasing difference in number of crosslinkers, the average of the difference in strain tends to increases, as expected. After 40 crosslinkers are removed, a statistically significant difference from the noise floor emerges. When examining strain maps for a particular sample depicting a magnitude of strain greater than 30\%, there is a bias towards positive strain for $N_c=2500$, $2490$, and $2940$ (Fig. 8(c)). Moreover, there appears to be a spatial localization of large negative strain springs that is somewhat conserved for the three cases. It is also interesting to note with more crosslinkers removed, there is an increase in the number of chromatin springs under large tension.

To make contact with more conventional chromatin configuration measures, Figs. 9(a)-(c) show the contact map, or adjacency matrix, when two particles are in contact or not with 2 chromatin particle diameters. If they are in contact, they are colored; otherwise, they are not. For $N_C=2500$, the particles in contact are denoted in red (Fig. 9(a)); for the removed crosslinker cases, the particles in contact are denoted in blue and are slightly larger squares(Figs. 9(b) and (c)). We can therefore overlay removed crosslinker cases with no crosslinker removed case, so we can see where there are differences. For $N_C=2500$ case, the two maps perfectly overlap, as they should, but for 10 and 60 crosslinkers removed, there are some purely red squares and some purely blue squares demonstrating the effect of eliminating crosslinkers. These regions should correlate with the differences in displacement maps shown in Figures 5 and 6 as differences in displacements can lead to changes in neighbors. Figure 9(d) demonstrates that the average number of contacts per particle decreases with fewer crosslinkers as expected.


\section{Discussion}

We posit a testable, multi-scale hypothesis for a difference in
brain organoid structures derived from human-derived pluripotent stem cells
and chimpanzee-derived pluripotent stem cells during the first ten days
of development. The hypothesis involves, ultimately, mechanical
perturbations of cell nuclei with human-derived pluripotent stem cells
demonstrating a different critical strain for particular regions of
chromatin organization as compared to chimpanzee-derived pluripotent
stem cells. The particular regions of chromatin organization are relevant to changes in gene expression of the ZEB2
transcription factor that can ultimately impact cell shape by way of
decreasing apical cell adhesion and increasing apical cell
constriction, as demonstrated previously~\cite{Lancaster2021}. 

While experimental confirmation awaits, we ask what insights can
computational modeling provide in terms of building such testable, multi-scale
hypotheses. To develop such insights, we argue that using cell-based,
computational modeling in terms of, for example, a three-dimensional vertex
model as presented in Ref. \cite{Zhang2022} is a reasonable starting point for an
organoid, particularly during the early stages of development. Such
models can ultimately provide accurate descriptions of cell
shape. To link the cell scale with the cell nuclear scale, we make the simplifying assumption that nuclear shape
tracks cell shape. And now that there exists a coarse-grained mechanical
model for a chromatin-containing cell nucleus that allows for
deformability and that recapitulates nuclear mechanics, correlated
chromatin motion, and changes in nuclear bleb initiation due to the
addition of $\alpha$-amanitin~\cite{banigan17,
Liu2021,Berg2023}, we can begin to study, at
some level, chromatin reorganization in response to mechanical
deformations. Other chromatin-based
models do not allow for nuclear deformability or account for the
lamina shell  
~\cite{saintillan18,Mahajan2022,Shi2018,Katava2022,Shi2023}.

Since our
focus is to look at structural differences between genetically-close
species whose chromatin organization is very similar at smaller
pair-wise scales but not necessarily at larger scales~\cite{Eres2019},
we determine 
how chromatin organization differs in response to mechanical
deformations between a reference state and a perturbed state with
fewer chromatin crosslinks for the same initial chromatin
configuration as a case study. We
denote the 
reference state as human state and the perturbed state as the
genetically-close relative. We find that as strain is increased on the
cell nucleus, it is only at larger strains that statistically
significant differences between the reference and the perturbed states
emerge.  Moreover, at larger strains of, for example, 40 \%, we find a
faster than linear scaling relation between increasingly less
genetically-close relatives and the average difference in
displacements between the reference (or human) chromatin state and the
genetically-close chromatin state.  Such a finding motivates the need
to quantify differences in spatiotemporal chromatin reorganization
across species just as comparative anatomy has done over centuries for
the brain and many other organs~\cite{Mota2015,Heuer2023}. Future work with enhanced statistics will make such scaling
relations more precise and, therefore, enhance the links between
tissue structure and chromatin organization.

To obtain more accurate models, on the other hand,
  one must go beyond the cell nuclear shape tracks cell shape
  assumption. For example, there is experimental evidence for
feedback between cell
shape and nuclear shape, with the nucleus releasing calcium should it
be compressed above a critical strain~\cite{Lomakin2020}.  As calcuim
promotes the vimentin filament polymerization, perhaps there is added
mechanical protection for the nucleus, in addition to actin
cytoskeletal mechanism~\cite{Lomakin2020}. We have not yet
explored such feedback. Moreover, since we are ultimately after a predictive model for changes in
chromatin configuration as a function of mechanical {\it and} chemical
perturbations, the cell nucleus model will require more detail such as
heterochromatin versus euchromatin and liquid-liquid phase separation
of chromatin crosslinkers~\cite{Strom2021} as well as a more accurate motor representation to
work towards predictive Hi-C in the presence of mechanical
perturbations, particularly as condensin II appears to determine genome
architecture across species~\cite{Hoencamp2021}. Efforts are already underway via HiCRes and HiCReg
that are rooted in libraries, or other means, but do not appear to focus on nuclear shape~\cite{Marchal2022,Shi2021}. Moreover, cell division plays an important
role and so understanding how chromatin reorganization during cell
division is important as well. 

As experimental scientists are able to obtain more detailed information at multiple
scales in living systems, it behooves the non-experimental scientists
to be able to stitch the scales together not just retroactively but
proactively, to be able to better understand the design
principles of life.  In other words, connecting the dots between genes
and tissues theoretically is becoming increasingly within our
reach. While here we focused on a multi-scale hypothesis for the
structure of brain organoids, one can obviously think more broadly to
organoids and tissues in general. Other modelers have begun to realize
the importance of multiscale mechanical modeling with a
two-dimensional vertex model containing rigid nuclei~\cite{Kim2022} and a 
powerful new three-dimensional mechanical model for cells that also includes
nuclei~\cite{Runser2024}. Both of these newer models do not yet include chromatin. 
Of course, certain types of questions regarding the structure of brain
organoids, for instance, do not require as detailed an
explicit framework containing chromatin and so we continue to work on answering them~\cite{Engstrom2018,Gandikota2021,Borzou2022}, though not
losing sight of the more detailed, multi-scale, computational modeling road that we are just
beginning to travel. We hope that many will join us along the way.

J.M.S. would like to acknowledge discussion with Orly Reiner, Amnon
Buxboim, Ken Kosik, Ardavan Borzou, Tjitse van der Molen, Alison Patteson, James Li, Alex Joyner, and Ed Banigan. TZ acknowledges the financial support from the NSFC/China (22303051). 
J.M.S. acknowledges financial
support from NSF-DMR-CMMT 1832002 and NSF-PHY-PoLS 2014192 and from an
Isaac Newton Award for Transformative Ideas during the COVID19
Pandemic from the DoD. M.A.L. is supported by the Medical Research
Council (MC\_UP\_1201/9). Some of the computations in this paper were run on the $\pi$ 2.0 and the Siyuan-1 cluster supported by the Center for High Performance Computing at Shanghai Jiao Tong University.

\bibliography{multiscale}

\end{document}